\documentclass[aip,jcp,reprint,superscriptaddress]{revtex4-1}

\usepackage{graphicx}
\usepackage{dcolumn}
\usepackage{subfigure}
\usepackage{multirow}
\usepackage{amssymb}
\usepackage{hyperref}
\usepackage{eucal}
\usepackage{upgreek}
\RequirePackage{lineno}

\newcommand{\llnl}{Lawrence Livermore National Laboratory, Livermore, CA 94550, USA}
\newcommand{\psu}{Department of Mechanical and Nuclear Engineering, The Pennsylvania State University, University Park, PA 16802, USA}

\newcommand{\pnnl}{Pacific Northwest National Laboratory, Richland, WA 99352, USA}
\newcommand{\be}{\begin{equation}}
\newcommand{\ee}{\end{equation}}

\newcommand{\iso}[2]{$^{#1}$#2}

\begin{document}

\title {Low-Energy ($<$10 keV) Electron Ionization and Recombination Model for a Liquid Argon Detector}



\author{M.~Foxe} \email[M. Foxe - ]{Michael.Foxe@pnnl.gov}\affiliation{\psu}\affiliation{\llnl}\affiliation{\pnnl}
\author{C.~Hagmann}\affiliation{\llnl}
\author{I.~Jovanovic} \affiliation{\psu}
\author{A.~Bernstein} \affiliation{\llnl}
\author{K.~Kazkaz} \affiliation{\llnl}
\author{V.~Mozin} \affiliation{\llnl}
\author{S.V.~Pereverzev} \affiliation{\llnl}
\author{S.~Sangiorgio} \affiliation{\llnl}
\author{P.~Sorensen} \affiliation{\llnl}



\date{\today}

\begin{abstract}

Detailed understanding of the ionization process in noble liquid detectors is important for their use in applications such as the search for dark matter and coherent elastic neutrino-nucleus scattering. The response of noble liquid detectors to low-energy ionization events is poorly understood at this time. We describe a new simulation tool which predicts the ionization yield from electronic energy deposits $(E < 10\, \rm keV)$ in liquid Ar, including the dependence of the yield on the applied electric drift field. The ionization signal produced in a liquid argon detector from \iso{37}{Ar} beta decay and \iso{55}{Fe} X-rays has been calculated using the new model. 

\end{abstract}


\maketitle

\section{Introduction}

Noble element detectors are of significant interest in the search for dark matter \cite{XENON10:2011, XENON100:2010, ZEPLIN2:2007, ZEPLIN3:2005, ArDM:2010, WARP:2005}. They exhibit both the sensitivity to detect small ionization signals and scalability to large detector masses. The state-of-the-art energy threshold in dual-phase Xe detector is the detection of a single electron \cite{Sorensen:2011}, while in a dual-phase Ar detector 270 eV electron recoil spectroscopy with single electron sensitivity has been demonstrated \cite{Sangiorgio:2013}. These detectors are usually calibrated with a variety of radiation sources, inducing either electron or nuclear recoils in the liquid. Electron recoils are used as the energy reference in modern dark matter detectors, and effort is made to calibrate nuclear energy depositions in terms of the reference electron energy depositions. In this paper we focus on the modeling of electron-induced ionization and recombination in liquid Ar. However striking similarities exist in the recombination physics for electron and nuclear recoils; the field dependent ionization yield from nuclear recoils is discussed separately \cite{Foxe:CNNSYield:2013}.

In order to minimize the detector energy threshold and the effect of local recombination, it is important to understand the combined effects of the externally applied electric drift field and electron-ion Coulomb interactions on the ionization yield. The empirical Thomas-Imel Box model \cite{Thomas:1987} has been used in the past to predict the magnitude of recombination in noble liquid detectors. In this model, a complete set of differential equations were written, with the diffusion and space charge terms subsequently neglected relative to electron drift and recombination. It is assumed that the recombination rate depends on the density of ions and electrons and the external field, but not on the Coulomb forces. Since the diffusion effects are ignored, all electron-ion pairs recombine for the case of zero external electric field. While the modified Thomas-Imel model \cite{Dahl:2009,NEST:2011} may fit the data well numerically, it does not provide insight into the physical processes. We aim to give a more complete picture of the ionization and thermalization process and provide an understanding of the spatial structure of the track.

In this paper, a model is presented which allows the electron ionization and recombination to be predicted at low energies. Starting with an initial fast electron, the resulting ionization cascade is simulated, followed by electron thermalization and drift under the influence of the electron and ion Coulomb field and the external drift field. We use the model to predict the response of a liquid Ar detector to two different low-energy ionizing radiation sources, \iso{37}{Ar} and \iso{55}{Fe}, at various electric drift fields. The model gives us a good qualitative picture of the structure of tracks created by incident projectiles. The model reveals that as the energy of the incident particle is reduced, the positive ion track length shortens while the size of the thermalized electron cloud stays constant. The recombination probability in an external field is a strong function of the total positive charge in the track core. However, our model does not exactly quantify  the parameters underlying the yield of electrons at these energies. Uncertainty comes from two sources. First, the model uses ionization cross sections  with binding energies that are characteristic of a gas, and does not completely account for interactions in the liquid state. Second, the thermalization process for electrons following ionization may not be fully accounted for in liquids in our (or other \cite{Wojcik:2003}) models.

\section{Electron Transport Model}

In order to predict the number of detectable electrons from events resulting in low energy deposition, an electron transport model has been developed. The electron positions are tracked using the electron transport algorithm, as described in Ref. \cite{Wojcik:2002}. This algorithm has previously successfully reproduced experimentally measured electron thermalization time \cite{Wojcik:2003,Mozumder:1995,Sowada:1982} and electron mobility in the electric field range of $\sim 1 - 10000$ V/cm \cite{Wojcik:2002}.The algorithm has also reproduced the electron escape probability as a function of ion separation \cite{Jaskolski:2011} obtained from the ICARUS experiment \cite{Amoruso:2004}.

A time step is chosen such that no more than one collision length is traversed by an electron within the full range of energies considered in the simulation. At each time step the forces on the electrons and ions due to Coulomb and external fields are calculated \cite{Jaskolski:2011}. Velocities and positions of electrons and ions are forward propagated to the subsequent time step of the simulation using the Verlet algorithm \cite{Jaskolski:2009}. Thermal motion of electrons is accounted for in the same way as in Ref. \cite{Jaskolski:2009}. The energy and momentum transfer cross sections for low-energy electron transport ($E\,<10\,{\rm eV}$) were also adopted from Ref. \cite{Wojcik:2002}. We neglect diffusion for positive Ar ions and use a constant mobility of $\upmu=0.6\times10^{-3}\text{ cm}^2/(\text{Vs})$ \cite{Henson:1964}.

\begin{figure}[tbp]
	\centering
	\includegraphics[width=\linewidth]{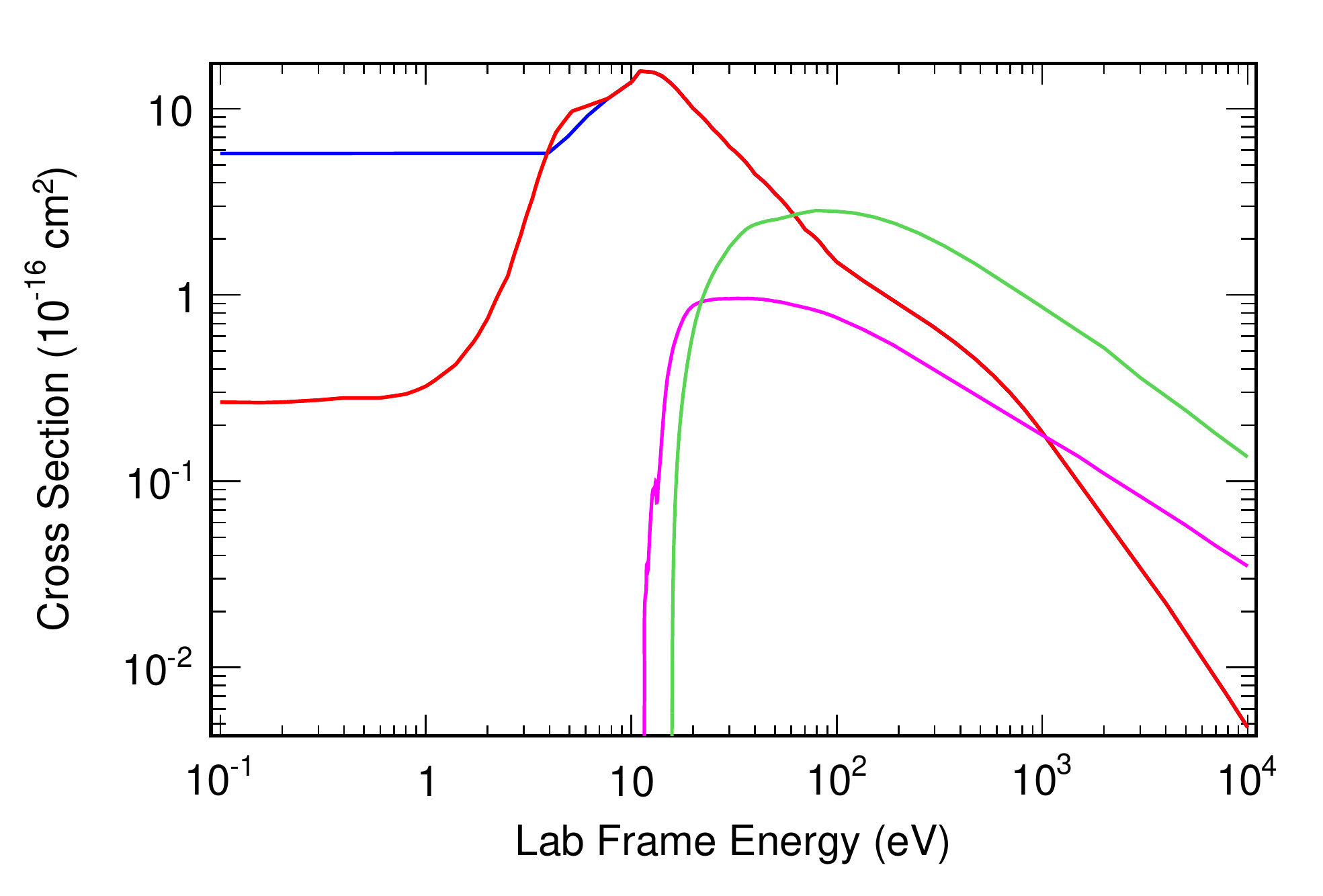}
	\caption[Cross section for electron-atom collisions]{Elastic and inelastic electron cross sections in liquid Ar. For E$<$10 eV, the elastic cross section comprises an energy transfer (blue) and a momentum transfer (red) component \cite{Wojcik:2002}. For E$>$10 eV, the momentum transfer cross section (red) is plotted \cite{lxcat:Phelps,Riley:1975}. Inelastic cross sections are divided into excitation and ionization. For excitation (magenta), a sum of 44 excitation cross sections is shown for E$<$1 keV\cite{lxcat:Biagi8.9,Biagi:1999}, and a single excitation cross section is used for E$>$1 keV \cite{lxcat:Phelps}. Similarly, for ionization (green), for E$<$1 keV, the cross section from Biagi is used \cite{lxcat:Biagi8.9,Biagi:1999}, and from Phelps for E$>$1 keV \cite{lxcat:Phelps}
}
	\label{fi:RecombinationXSec}
\end{figure}

The electron energy range is extended by including elastic, excitation, and ionization cross sections for Ar up to $E$ = 10 keV. The Ar cross sections for $E\, <$ 1000 eV originate from Biagi \cite{Biagi:1999,lxcat:Biagi8.9} and include ~ 44 individual atomic excitation levels. At higher energies, the data from Phelps \cite{lxcat:Phelps} with a single effective excitation cross section were used. For elastic electron scattering, we used Biagi's momentum transfer cross section up to $E$ = 100 eV. Above that value we adopted the detailed angular scattering distributions by Fink \cite{Fink:1970} up to $E$ = 1 keV and Riley's distributions above 1 keV \cite{Riley:1975}. The table energy is sampled using statistical interpolation, where the table energy on the low and high side of the current energy $E$ is sampled with probability $(E_{hi}-E)/(E_{hi}-E_{lo})$ and $(E-E_{lo})/(E_{hi}-E_{lo})$, respectively. The cross sections for each interaction are shown in Figure \ref{fi:RecombinationXSec}.

The electron collisions are sampled using the null collision method, with a real electron interaction occurring with a probability
\be
Prob = \frac{v\ \sigma_{tot}(v)}{K_{\text{max}}},
\label{eq:ElecTransColl}
\ee
where $v$ is the electron velocity, $\sigma_{tot}(v)$ is the total cross section for interaction, and $K_{\text{max}} \equiv {\rm max}( v\ \sigma_{tot}(v)  )$ is the maximum collision rate in the considered electron energy range.

If a real collision occurs, another sampling is performed to determine if the resulting collision is elastic or inelastic. For elastic collisions at energies $<~ 10$ eV, a choice must be made between momentum transfer collisions with cross section $\sigma_p(v)$ and energy transfer collisions represented by $\sigma_e(v)$. The former process consists of isotropic electron-Ar scattering in the center-of-mass frame with change in electron direction and energy, whereas the latter process results in a change of the electron energy only. These two cross sections are tuned to reproduce the measured electron drift velocities in liquid Ar.
 
\begin{figure}[tbp]
	\centering
	\includegraphics[width=\linewidth]{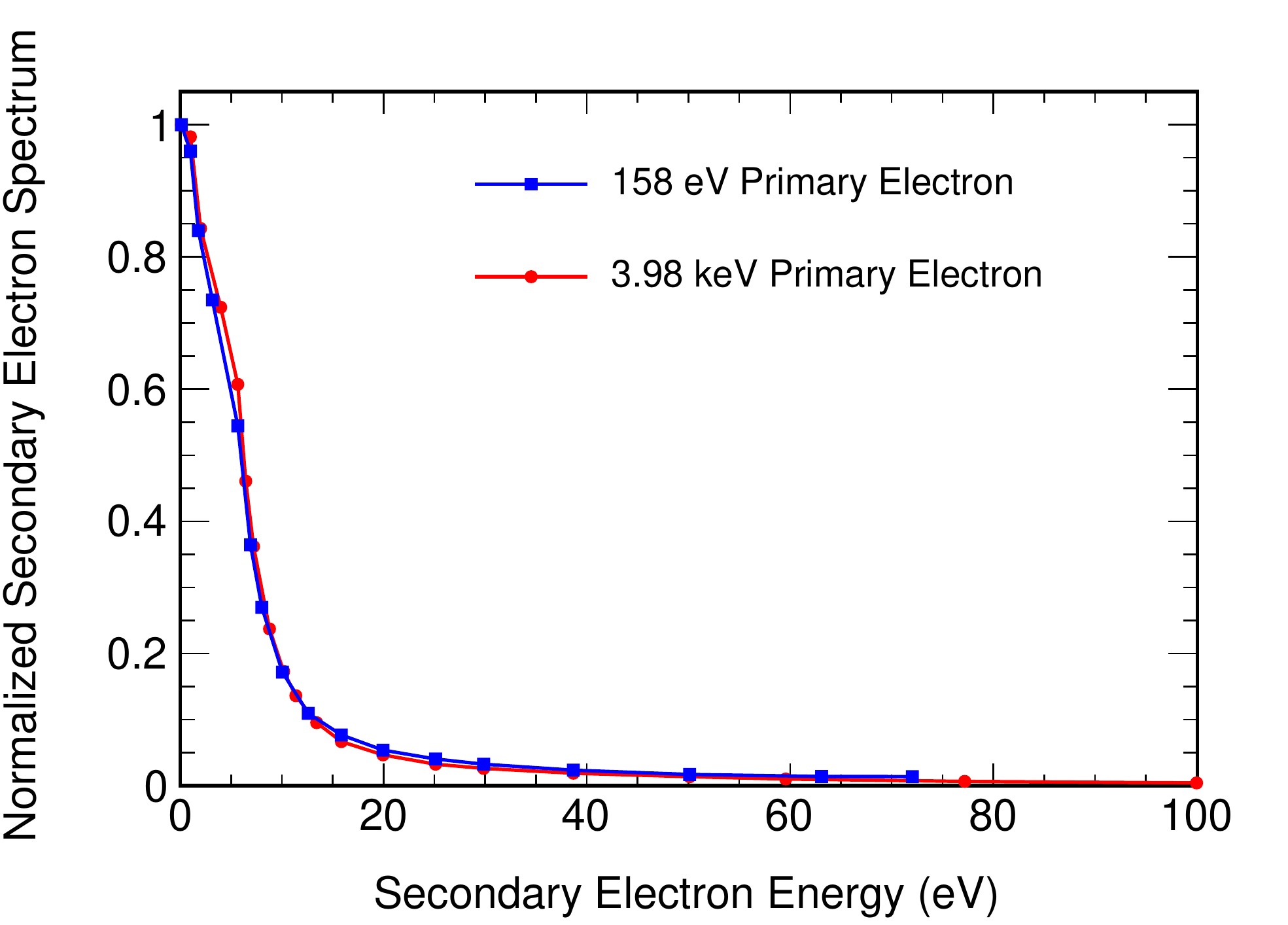}
	\caption[]{Normalized secondary electron spectra for the M3-shell in Ar as tabulated in the EEDL database \cite{EEDL:1991}.}
	\label{fi:EEDL}
\end{figure}

If an atomic excitation is sampled, the excitation energy of the atomic level is subtracted from the electron energy, while the electron direction is kept unchanged. If an ionization occurs, the emission angle of the secondary electron is randomly sampled, with the primary electron's direction kept constant. The secondary electron energy is sampled from probability distributions in the EEDL database. At keV energies, the total ionization cross section is dominated by the contribution from the M3 shell. As shown in Figure \ref{fi:EEDL}, the spectra vary very little at higher energies and we use the plotted distribution for all energies $>158$ eV. At lower energies, a linearly interpolated distribution with 0 eV at threshold and the EEDL values at $E=158$ eV is used. For all cases, the electron energy was sampled using the rejection method.

The model is used to calculate the expected electron production in \iso{37}{Ar} and \iso{55}{Fe} decays. \iso{37}{Ar} decays via electron capture followed by X-ray or Auger electron emission; the electron capture branching ratios are shown in Table \ref{tb:Ar37DecayScheme}. \iso{55}{Fe} decays via X-ray emission, with the emission probabilities shown in Table \ref{tb:Fe55XRay}. The \iso{55}{Fe} source was modeled with a single 5.9 keV X-ray absorbed in the Ar. Approximately 90 \% of the X-rays create a K-shell vacancy in Ar \cite{Cul1991,CulData:1991}; the photo-absorption in outer shells and Compton scattering are neglected. The photoelectron energy is $E_{p.e.} = 5.9\,{\rm keV} - E{_b} = 2.77\,\rm{keV}$, where $E_b$ is the K-shell binding energy, and the remaining energy is emitted as Auger electrons and X-rays. The relaxation cascade was sampled from the EADL database \cite{Cul1991,CulData:1991}, which predicts that on average 3.3 Auger electrons are emitted with a total energy of 3.1 keV. The probability of X-ray fluorescence is very small and X-ray fluorescence is neglected in the simulation. All Auger electrons and photoelectrons are emitted isotropically with the number of positive ions balancing the free electrons.

The decay of \iso{37}{Ar} leads to either a K-shell or a L1-shell (the lowest L-shell) vacancy in the \iso{37}{Cl} daughter. Its atomic relaxation is treated in a fashion similar to \iso{55}{Fe} and produces on average 3.3 (2.0) Auger electrons carrying away 2.74 (0.22) keV of energy for K (L1) shell vacancies. A 1.0 keV electron was also simulated to bridge the energy gap between the \iso{37}{Ar} K-shell and L1-shell events.

\begin{table}[tbp]
\caption[Electron capture decay branching ratios for \iso{37}{Ar}.]{Electron capture decay branching ratios for \iso{37}{Ar} (\iso{37}{Ar} decays 100\% via electron capture).\cite{Barsanov:2007}}
\begin{center}
\begin{tabular}{ccc}
\hline
Energy [keV] & Branching Ratio& Decay Mode \\
\hline
2.8224 & $0.9017 \pm 0.0024$ & K-shell Capture \\
0.2702 & $0.0890 \pm 0.0027$ & L-shell Capture \\
0.0175 & $0.0093^{+0.0006}_{-0.0004}$ & M-shell Capture \\
\hline
\end{tabular}
\end{center}
\label{tb:Ar37DecayScheme}
\end{table}

\begin{table}[tbp]
\caption[X-ray emission energies for the \iso{55}{Fe} source]{The most intense X-rays from \iso{55}{Fe}, reproduced from the Table of Isotopes~\cite{LBL2007}. The most intense X-ray not included in this Table has a relative intensity of 0.28\%, approximately 100 times smaller than the intensity of the combined 5.888 and 5.899 keV X-rays \cite{Kazkaz:2010}.}
\begin{center}
\begin{tabular}{cc}
\hline
Energy (keV) & Relative Intensity (\%) \\
\hline
5.888 & 8.5 \\
5.899 & 16.98 \\
6.490 & 1.01 \\
6.490 & 1.98 \\
\hline
\end{tabular}
\end{center}
\label{tb:Fe55XRay}
\end{table}

After the initial emission of the Auger and photoelectrons, all of the electrons are tracked using the electron transport model, generating ionization and excitation events as the electrons slow down and eventually thermalize. 

\section{Electron Transport Model Results}

\begin{figure}[tbp]
	\centering
	\includegraphics[width=\linewidth]{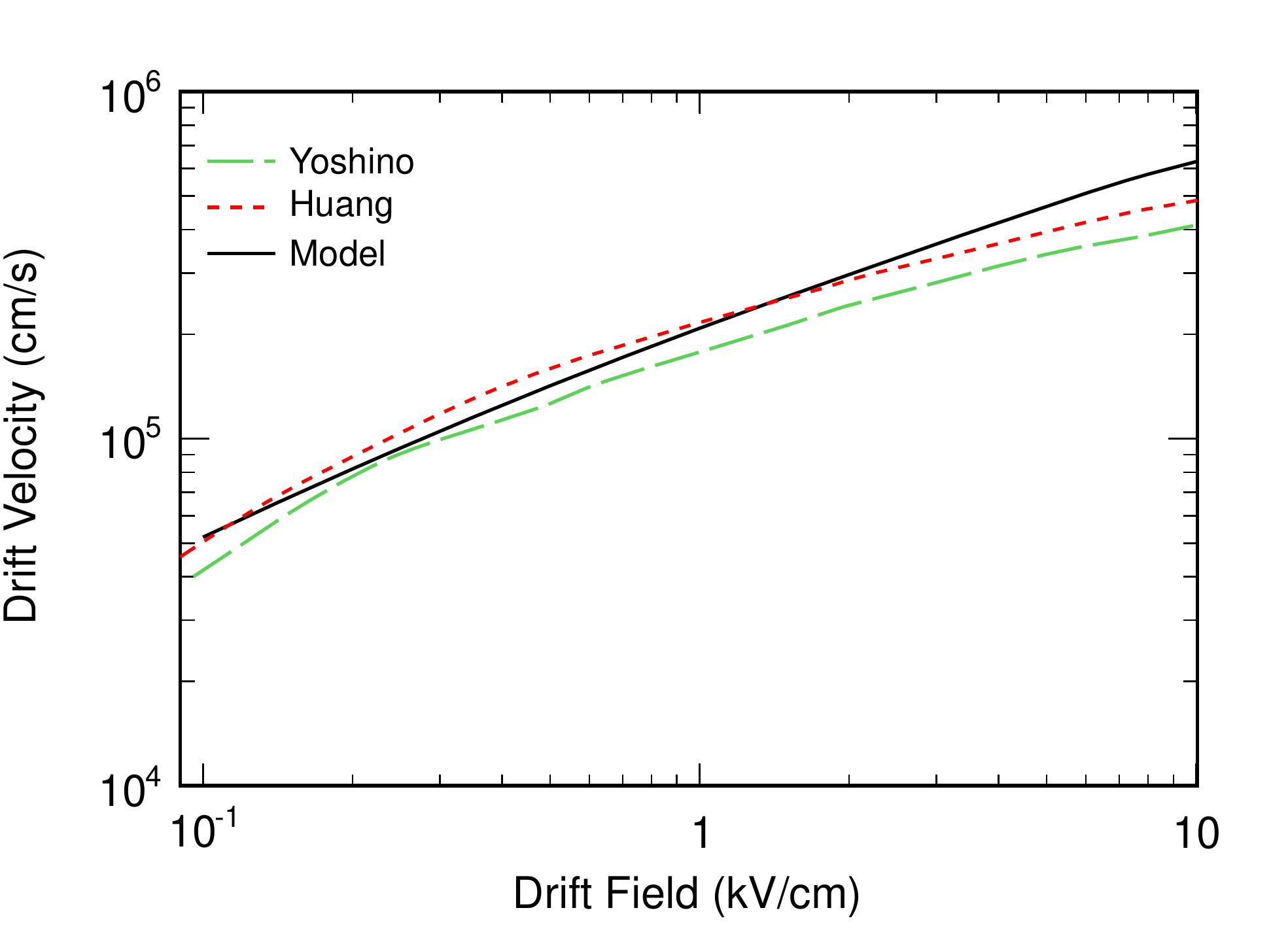}
	\caption[]{Calculated electron drift velocities in liquid Ar as a function of electric field. The model (black) reproduces the curve obtained by Wojcik \cite{Wojcik:2002} (not shown), and matches with experimental data by Yoshino \cite{Yoshino:1976} (green) and Huang \cite{Huang:1981} (red).}
	\label{fi:DriftVelCheck}
\end{figure}

As part of the code validation, the electron drift speed as a function of electric field was calculated. Our results shown in Figure \ref{fi:DriftVelCheck} closely reproduce the results obtained by Wojcik \cite{Wojcik:2002}. Also shown are experimental data by Yoshino \cite{Yoshino:1976} and Huang \cite{Huang:1981}. Except at the highest fields, the agreement with the model is quite good.

\begin{figure}[tbp]
	\centering
	\includegraphics[width=\linewidth]{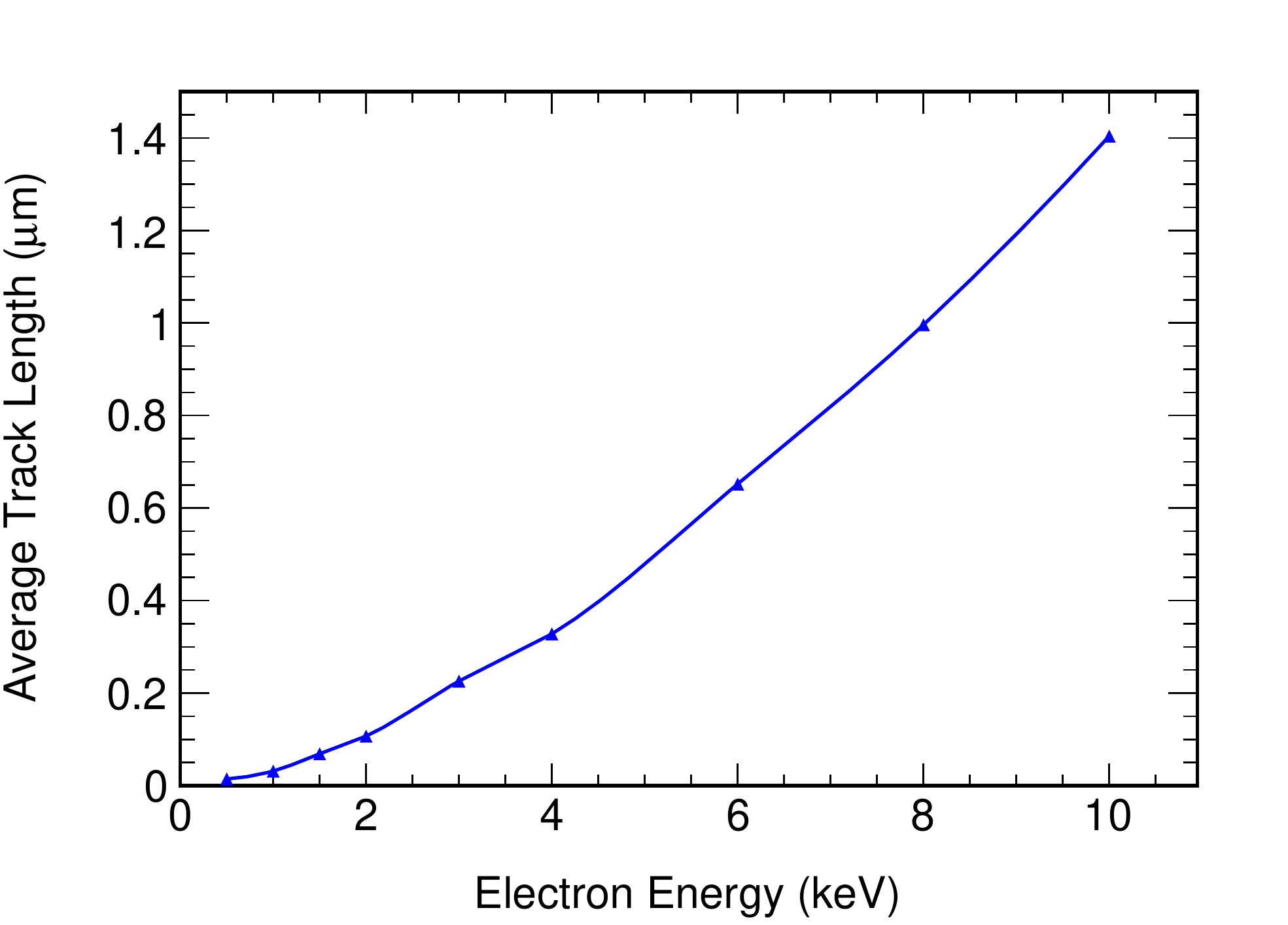}
	\caption[]{Average ionization track lengths calculated in the simulation are short compared to the electron thermalization length of 2.6 $\upmu$m reported by Wojcik \cite{Wojcik:2003}. Electrons generated from \iso{37}{Ar} and \iso{55}{Fe} are $<\sim1/10$ of the thermalization length.}
	\label{fi:TrackLength}
\end{figure}

We also studied the average ionization track length for a single fast electron as a function of energy $E\,<$ 10 keV. As shown in Figure \ref{fi:TrackLength}, the ionization track length is short compared to the electron thermalization length $\sim$ 2.6 $\mu$m reported by Wojcik \cite{Wojcik:2003}. This corresponds to the picture of a short positive ion track surrounded by a spherical electron cloud at the end of an electron thermalization. If an external field is present, the electrons are accelerated away from the ions, but due to the attractive Coulomb force, some electrons become trapped and recombine with positive ions. We adopted Wojcik's criterion for recombination, i.e. $E\,<\,E_{crit}\, =$ 1 eV and $r_{electron-ion}\,<\,r_{crit}\,=\,1.3\,$nm \cite{Jaskolski:2009}. Following recombination, the electron and ion are removed from the simulation.

\begin{figure}[tbp]
	\centering
	\includegraphics[width=\linewidth]{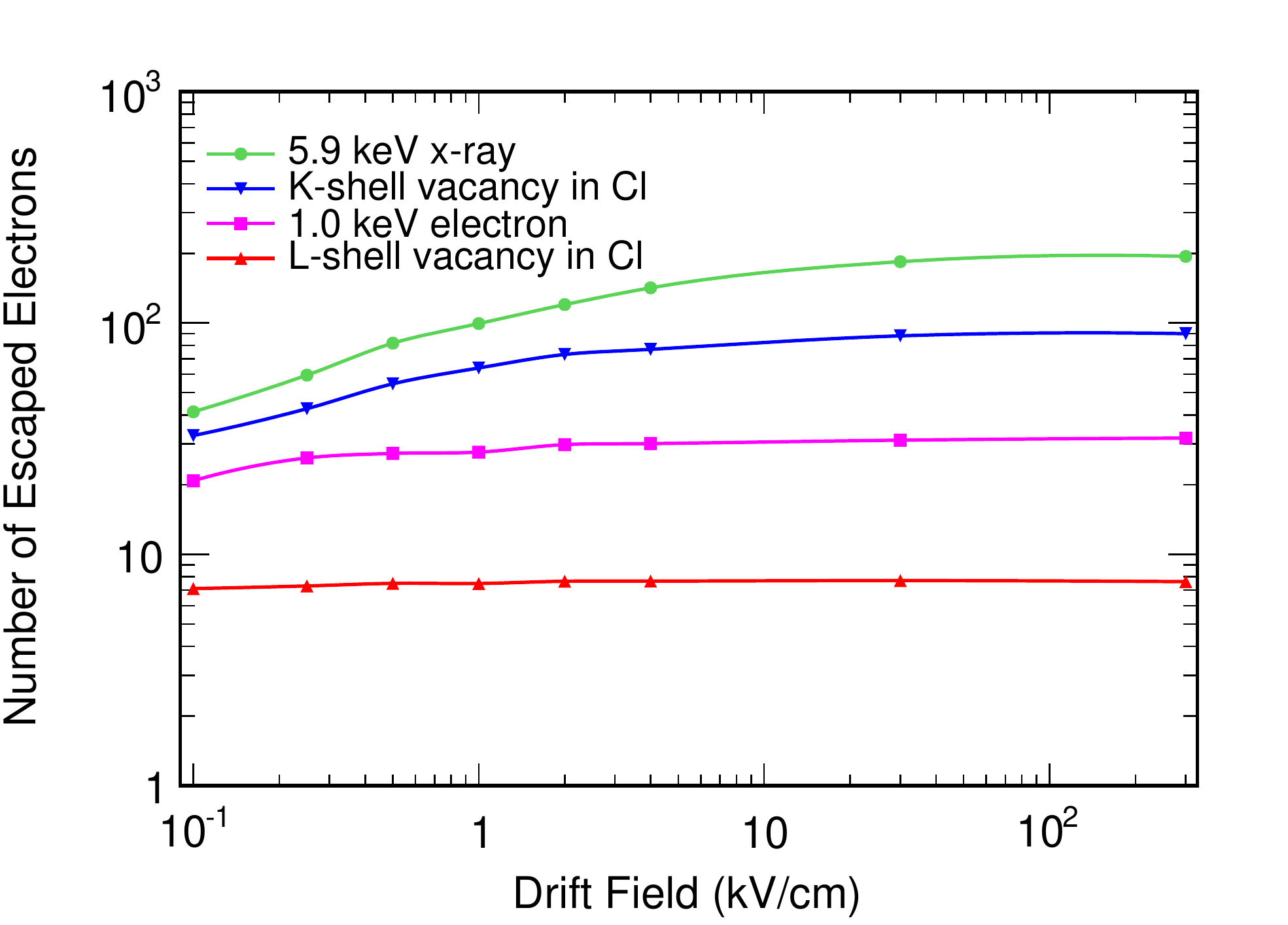}
	\caption[Calculated electron yield from \iso{37}{Ar} L-shell as a function of drift electric field]{Calculated number of electrons that escape the ion cloud as a function of electric field for the \iso{37}{Ar} L1-shell peak (red), 1.0 keV electron (magenta), \iso{37}{Ar} K-shell peak (blue), and \iso{55}{Fe} X-ray peak (green). The low-energy deposits associated with the \iso{37}{Ar} L1-shell and, to a lesser extent, the 1.0 keV electrons exhibit no field dependence for drift fields $>0.1$ kV/cm. For the higher energy events associated with the \iso{37}{Ar} K-shell and \iso{55}{Fe} X-rays, the electron yield depends strongly on the applied electric field.}
	\label{fi:RecombCalc}
\end{figure}

Simulations of ionization tracks from \iso{55}{Fe} X-rays and \iso{37}{Ar} electron captures were performed for the electric field range of 0.1 - 300 kV/cm. Each calculation was performed sufficiently long to allow for electron thermalization and the separation of the electron cloud from the ion track. For drift fields $<$ 1 kV/cm simulation was conducted over $\sim$ 13 ns; for larger fields a simulation time of $\sim$ 5 ns was adequate. For each drift field value and electron source type, 20-100 histories with different random number seeds were performed and averaged. The number of electrons that escape from the initial ion cloud is shown in Figure \ref{fi:RecombCalc}. Due to the low amount of ionization generated from \iso{37}{Ar} L-shell electron capture, the electrons are capable of fully escaping the ion cloud even at low electric fields. Simulations of higher energy events from the K-shell electron capture of \iso{37}{Ar} and the decay of \iso{55}{Fe} exhibit a clear dependence on the electric field, with the electron escape probability approximately doubling between 0.5 kV/cm and 4 kV/cm in liquid Ar. Additionally, it can be seen that the higher the deposited energy, the greater is the dependence of the electron escape probability on the electric field. The electrons that escape the ion cloud can potentially be detected, if they are not lost to capture on impurities within the detector.

The electron transport model produces three quantities: the number of ionizations, excitations, and the number of electrons that escape from the ion cloud. The total number of ionizations and excitations to occur during the thermalization process is shown in Table \ref{tb:TotalChargesCalc}. Calculations were made using the electron transport model at 300 kV/cm to obtain the asymptotically large drift field W-value, or the average energy required to generate an electron which escapes the initial ion cloud. The calculated asymptotically large drift field $W$-values for \iso{37}{Ar} and \iso{55}{Fe} are given in Table \ref{tb:InfWValuesCalc}. 

\begin{table}[tbp]
\caption[Average number of electrons and photons generated]{The electron transport code calculates the total number of electrons and photons created during the thermalization process of the initial high energy electrons for \iso{37}{Ar} and \iso{55}{Fe}. Errors given correspond to a 1-sigma variation, based on 1000 simulations.}
\begin{center}
\begin{tabular}{cccc}
\hline
Isotope & Ionizations & Excitations & $N_{\text{exc}}/N_{\text{ion}}$\\
\hline
\iso{37}{Ar} L-shell & $7.6 \pm 1.2$ & $4.3 \pm 1.8$& $0.57 \pm 0.25$ \\
\iso{37}{Ar} K-shell  & $89.9 \pm 4.2$ & $41.8 \pm 5.7$& $ 0.47 \pm 0.07$ \\
\iso{55}{Fe}  & $193.9 \pm 6.6$ & $87.2 \pm 8.0$& $ 0.45 \pm 0.04$ \\
\hline
\end{tabular}
\end{center}
\label{tb:TotalChargesCalc}
\end{table}
\begin{table}[tbp]
\caption[Calculated asymptotically large drift field W-Values for LAr]{The asymptotically large drift field $W$-values for liquid argon were calculated using the electron transport code with an electric field set to 300 kV/cm. Errors given correspond to a 1-sigma variation, based on 1000 simulations.}
\begin{center}
\begin{tabular}{ccc}
\hline
Decay Isotope & Energy Available (keV) & $W$-Value (eV) \\
\hline
\iso{37}{Ar} L-shell & 0.22 & $28.9 \pm 4.4$\\
\iso{37}{Ar} K-shell & 2.74 & $30.5 \pm 1.4$\\
\iso{55}{Fe} & 5.9 & $30.4 \pm 1.0$\\
\hline
\end{tabular}
\end{center}
\label{tb:InfWValuesCalc}
\end{table}

\section{Conclusions}
The previously published electron transport algorithm \cite{Wojcik:2002} has been extended to model low-energy $(E\,<\,10 \,{\rm keV})$ electron-induced ionization in liquid Ar. The formation and evolution of the electron and ion cloud is treated in detail, including local recombination. The simulation predicts very little recombination in typical drift fields of 1 kV/cm for electron energy depositions $E\,< $ 1 keV, corresponding to $\sim$ 30 ionization electrons. For larger energy depositions, the Coulomb forces between ions and electrons tend to increase the local recombination rate, especially in small drift fields. For example, the model predicts a loss of $\sim$ 50\% of initial electrons to recombination from a \iso{55}{Fe} X-ray in a drift field of 1 kV/cm compared to the asymptotically large drift field yield. We plan to apply our model to calculate the fraction of electrons undergoing local recombination in low-energy nuclear recoils in liquid Ar, leading to more precise predictions of ionization signals in low-threshold neutrino and dark matter detectors.
  


\begin{acknowledgments}
This work was performed under the auspices of the U.S. Department of Energy by the Lawrence Livermore National Laboratory under Contract DE-AC52-07NA27344. A portion of M. Foxe's research was performed under the Nuclear Forensics Graduate Fellowship Program, which is sponsored by the U.S. Department of Homeland Security, Domestic Nuclear Detection Office and the U.S. Department of Defense, Defense Threat Reduction Agency.  PNNL-SA-100217
\end{acknowledgments}


\begin{thebibliography}{33}%
\makeatletter
\providecommand \@ifxundefined [1]{%
 \@ifx{#1\undefined}
}%
\providecommand \@ifnum [1]{%
 \ifnum #1\expandafter \@firstoftwo
 \else \expandafter \@secondoftwo
 \fi
}%
\providecommand \@ifx [1]{%
 \ifx #1\expandafter \@firstoftwo
 \else \expandafter \@secondoftwo
 \fi
}%
\providecommand \natexlab [1]{#1}%
\providecommand \enquote  [1]{``#1''}%
\providecommand \bibnamefont  [1]{#1}%
\providecommand \bibfnamefont [1]{#1}%
\providecommand \citenamefont [1]{#1}%
\providecommand \href@noop [0]{\@secondoftwo}%
\providecommand \href [0]{\begingroup \@sanitize@url \@href}%
\providecommand \@href[1]{\@@startlink{#1}\@@href}%
\providecommand \@@href[1]{\endgroup#1\@@endlink}%
\providecommand \@sanitize@url [0]{\catcode `\\12\catcode `\$12\catcode
  `\&12\catcode `\#12\catcode `\^12\catcode `\_12\catcode `\%12\relax}%
\providecommand \@@startlink[1]{}%
\providecommand \@@endlink[0]{}%
\providecommand \url  [0]{\begingroup\@sanitize@url \@url }%
\providecommand \@url [1]{\endgroup\@href {#1}{\urlprefix }}%
\providecommand \urlprefix  [0]{URL }%
\providecommand \Eprint [0]{\href }%
\providecommand \doibase [0]{http://dx.doi.org/}%
\providecommand \selectlanguage [0]{\@gobble}%
\providecommand \bibinfo  [0]{\@secondoftwo}%
\providecommand \bibfield  [0]{\@secondoftwo}%
\providecommand \translation [1]{[#1]}%
\providecommand \BibitemOpen [0]{}%
\providecommand \bibitemStop [0]{}%
\providecommand \bibitemNoStop [0]{.\EOS\space}%
\providecommand \EOS [0]{\spacefactor3000\relax}%
\providecommand \BibitemShut  [1]{\csname bibitem#1\endcsname}%
\let\auto@bib@innerbib\@empty
\bibitem [{\citenamefont {{XENON10 Collaboration}}(2011)}]{XENON10:2011}%
  \BibitemOpen
  \bibfield  {author} {\bibinfo {author} {\bibnamefont {{XENON10
  Collaboration}}},\ }\href {\doibase 10.1016/j.astropartphys.2011.01.006}
  {\bibfield  {journal} {\bibinfo  {journal} {Astroparticle Physics}\ }\textbf
  {\bibinfo {volume} {34}},\ \bibinfo {pages} {679} (\bibinfo {year}
  {2011})}\BibitemShut {NoStop}%
\bibitem [{\citenamefont {{XENON100 Collaboration}}(2010)}]{XENON100:2010}%
  \BibitemOpen
  \bibfield  {author} {\bibinfo {author} {\bibnamefont {{XENON100
  Collaboration}}},\ }\href {\doibase 10.1103/PhysRevLett.105.131302}
  {\bibfield  {journal} {\bibinfo  {journal} {Physical Review Letters}\
  }\textbf {\bibinfo {volume} {105}},\ \bibinfo {pages} {131302} (\bibinfo
  {year} {2010})}\BibitemShut {NoStop}%
\bibitem [{\citenamefont {{ZEPLIN-II Collaboration}}(2007)}]{ZEPLIN2:2007}%
  \BibitemOpen
  \bibfield  {author} {\bibinfo {author} {\bibnamefont {{ZEPLIN-II
  Collaboration}}},\ }\href {\doibase 10.1016/j.astropartphys.2007.06.002}
  {\bibfield  {journal} {\bibinfo  {journal} {Astroparticle Physics}\ }\textbf
  {\bibinfo {volume} {28}},\ \bibinfo {pages} {287} (\bibinfo {year}
  {2007})}\BibitemShut {NoStop}%
\bibitem [{\citenamefont {{ZEPLIN Collaboration}}(2005)}]{ZEPLIN3:2005}%
  \BibitemOpen
  \bibfield  {author} {\bibinfo {author} {\bibnamefont {{ZEPLIN
  Collaboration}}},\ }\href {\doibase 10.1016/j.newar.2005.01.018} {\bibfield
  {journal} {\bibinfo  {journal} {New Astronomy Reviews}\ }\textbf {\bibinfo
  {volume} {49}},\ \bibinfo {pages} {277} (\bibinfo {year} {2005})}\BibitemShut
  {NoStop}%
\bibitem [{\citenamefont {{ArDM Collaboration}}(2008)}]{ArDM:2010}%
  \BibitemOpen
  \bibfield  {author} {\bibinfo {author} {\bibnamefont {{ArDM Collaboration}}}\
  }(\bibinfo {year} {2008})\BibitemShut {NoStop}%
\bibitem [{\citenamefont {Brunetti}\ \emph {et~al.}(2005)\citenamefont
  {Brunetti}, \citenamefont {Calligarich}, \citenamefont {Cambiaghi},
  \citenamefont {Carbonara}, \citenamefont {Cocco}, \citenamefont {Vecchi},
  \citenamefont {Dolfini}, \citenamefont {Ereditato}, \citenamefont {Fiorillo},
  \citenamefont {Grandi}, \citenamefont {Mangano}, \citenamefont {Menegolli},
  \citenamefont {Montanari}, \citenamefont {Prata}, \citenamefont {Rappoldi},
  \citenamefont {Raselli}, \citenamefont {Roncadelli}, \citenamefont
  {Rossella}, \citenamefont {Rubbia}, \citenamefont {Santorelli},\ and\
  \citenamefont {Vignoli}}]{WARP:2005}%
  \BibitemOpen
  \bibfield  {author} {\bibinfo {author} {\bibfnamefont {R.}~\bibnamefont
  {Brunetti}}, \bibinfo {author} {\bibfnamefont {E.}~\bibnamefont
  {Calligarich}}, \bibinfo {author} {\bibfnamefont {M.}~\bibnamefont
  {Cambiaghi}}, \bibinfo {author} {\bibfnamefont {F.}~\bibnamefont
  {Carbonara}}, \bibinfo {author} {\bibfnamefont {A.}~\bibnamefont {Cocco}},
  \bibinfo {author} {\bibfnamefont {C.~D.}\ \bibnamefont {Vecchi}}, \bibinfo
  {author} {\bibfnamefont {R.}~\bibnamefont {Dolfini}}, \bibinfo {author}
  {\bibfnamefont {A.}~\bibnamefont {Ereditato}}, \bibinfo {author}
  {\bibfnamefont {G.}~\bibnamefont {Fiorillo}}, \bibinfo {author}
  {\bibfnamefont {L.}~\bibnamefont {Grandi}}, \bibinfo {author} {\bibfnamefont
  {G.}~\bibnamefont {Mangano}}, \bibinfo {author} {\bibfnamefont
  {A.}~\bibnamefont {Menegolli}}, \bibinfo {author} {\bibfnamefont
  {C.}~\bibnamefont {Montanari}}, \bibinfo {author} {\bibfnamefont
  {M.}~\bibnamefont {Prata}}, \bibinfo {author} {\bibfnamefont
  {A.}~\bibnamefont {Rappoldi}}, \bibinfo {author} {\bibfnamefont
  {G.}~\bibnamefont {Raselli}}, \bibinfo {author} {\bibfnamefont
  {M.}~\bibnamefont {Roncadelli}}, \bibinfo {author} {\bibfnamefont
  {M.}~\bibnamefont {Rossella}}, \bibinfo {author} {\bibfnamefont
  {C.}~\bibnamefont {Rubbia}}, \bibinfo {author} {\bibfnamefont
  {R.}~\bibnamefont {Santorelli}}, \ and\ \bibinfo {author} {\bibfnamefont
  {C.}~\bibnamefont {Vignoli}},\ }\href {\doibase 10.1016/j.newar.2005.01.017}
  {\bibfield  {journal} {\bibinfo  {journal} {New Astronomy Reviews}\ }\textbf
  {\bibinfo {volume} {49}},\ \bibinfo {pages} {265} (\bibinfo {year}
  {2005})}\BibitemShut {NoStop}%
\bibitem [{\citenamefont {Sorensen}\ and\ \citenamefont
  {Dahl}(2011)}]{Sorensen:2011}%
  \BibitemOpen
  \bibfield  {author} {\bibinfo {author} {\bibfnamefont {P.}~\bibnamefont
  {Sorensen}}\ and\ \bibinfo {author} {\bibfnamefont {C.~E.}\ \bibnamefont
  {Dahl}},\ }\href {\doibase 10.1103/PhysRevD.83.063501} {\bibfield  {journal}
  {\bibinfo  {journal} {Physical Review D}\ }\textbf {\bibinfo {volume} {83}},\
  \bibinfo {pages} {063501} (\bibinfo {year} {2011})}\BibitemShut {NoStop}%
\bibitem [{\citenamefont {Sangiorgio}\ \emph {et~al.}(2013)\citenamefont
  {Sangiorgio}, \citenamefont {Joshi}, \citenamefont {Bernstein}, \citenamefont
  {Coleman}, \citenamefont {Foxe}, \citenamefont {Hagmann}, \citenamefont
  {Jovanovic}, \citenamefont {Kazkaz}, \citenamefont {Mavrokoridis},
  \citenamefont {Mozin}, \citenamefont {Pereverzev},\ and\ \citenamefont
  {Sorensen}}]{Sangiorgio:2013}%
  \BibitemOpen
  \bibfield  {author} {\bibinfo {author} {\bibfnamefont {S.}~\bibnamefont
  {Sangiorgio}}, \bibinfo {author} {\bibfnamefont {T.}~\bibnamefont {Joshi}},
  \bibinfo {author} {\bibfnamefont {A.}~\bibnamefont {Bernstein}}, \bibinfo
  {author} {\bibfnamefont {J.}~\bibnamefont {Coleman}}, \bibinfo {author}
  {\bibfnamefont {M.}~\bibnamefont {Foxe}}, \bibinfo {author} {\bibfnamefont
  {C.}~\bibnamefont {Hagmann}}, \bibinfo {author} {\bibfnamefont
  {I.}~\bibnamefont {Jovanovic}}, \bibinfo {author} {\bibfnamefont
  {K.}~\bibnamefont {Kazkaz}}, \bibinfo {author} {\bibfnamefont
  {K.}~\bibnamefont {Mavrokoridis}}, \bibinfo {author} {\bibfnamefont
  {V.}~\bibnamefont {Mozin}}, \bibinfo {author} {\bibfnamefont
  {S.}~\bibnamefont {Pereverzev}}, \ and\ \bibinfo {author} {\bibfnamefont
  {P.}~\bibnamefont {Sorensen}},\ }\href {\doibase
  http://dx.doi.org/10.1016/j.nima.2013.06.061} {\bibfield  {journal} {\bibinfo
   {journal} {Nuclear Instruments and Methods in Physics Research Section A:
  Accelerators, Spectrometers, Detectors and Associated Equipment}\ }\textbf
  {\bibinfo {volume} {728}},\ \bibinfo {pages} {69 } (\bibinfo {year}
  {2013})}\BibitemShut {NoStop}%
\bibitem [{\citenamefont {Foxe}\ \emph {et~al.}()\citenamefont {Foxe},
  \citenamefont {Hagmann}, \citenamefont {Jovanovic}, \citenamefont
  {Bernstein}, \citenamefont {Joshi}, \citenamefont {Kazkaz}, \citenamefont
  {Mozin}, \citenamefont {Pereverzev}, \citenamefont {Sangiorgio},\ and\
  \citenamefont {Sorensen}}]{Foxe:CNNSYield:2013}%
  \BibitemOpen
  \bibfield  {author} {\bibinfo {author} {\bibfnamefont {M.}~\bibnamefont
  {Foxe}}, \bibinfo {author} {\bibfnamefont {C.}~\bibnamefont {Hagmann}},
  \bibinfo {author} {\bibfnamefont {I.}~\bibnamefont {Jovanovic}}, \bibinfo
  {author} {\bibfnamefont {A.}~\bibnamefont {Bernstein}}, \bibinfo {author}
  {\bibfnamefont {T.}~\bibnamefont {Joshi}}, \bibinfo {author} {\bibfnamefont
  {K.}~\bibnamefont {Kazkaz}}, \bibinfo {author} {\bibfnamefont
  {V.}~\bibnamefont {Mozin}}, \bibinfo {author} {\bibfnamefont
  {S.}~\bibnamefont {Pereverzev}}, \bibinfo {author} {\bibfnamefont
  {S.}~\bibnamefont {Sangiorgio}}, \ and\ \bibinfo {author} {\bibfnamefont
  {P.}~\bibnamefont {Sorensen}},\ }\href@noop {} {\bibfield  {journal}
  {\bibinfo  {journal} {Astroparticle Physics}\ }\textbf {\bibinfo {volume} {in
  preparation}}}\BibitemShut {NoStop}%
\bibitem [{\citenamefont {Thomas}\ and\ \citenamefont
  {Imel}(1987)}]{Thomas:1987}%
  \BibitemOpen
  \bibfield  {author} {\bibinfo {author} {\bibfnamefont {J.}~\bibnamefont
  {Thomas}}\ and\ \bibinfo {author} {\bibfnamefont {D.~A.}\ \bibnamefont
  {Imel}},\ }\href {\doibase 10.1103/PhysRevA.36.614} {\bibfield  {journal}
  {\bibinfo  {journal} {Physical Review A}\ }\textbf {\bibinfo {volume} {36}},\
  \bibinfo {pages} {614} (\bibinfo {year} {1987})}\BibitemShut {NoStop}%
\bibitem [{\citenamefont {Dahl}(2009)}]{Dahl:2009}%
  \BibitemOpen
  \bibfield  {author} {\bibinfo {author} {\bibfnamefont {E.}~\bibnamefont
  {Dahl}},\ }\emph {\bibinfo {title} {The Physics of Background Discrimination
  in Liquid Xenon, and First Results from Xenon10 in the Hunt for {WIMP} Dark
  Matter}},\ \href@noop {} {\bibinfo {type} {{Ph.D.}}},\ \bibinfo  {school}
  {Princeton University} (\bibinfo {year} {2009})\BibitemShut {NoStop}%
\bibitem [{\citenamefont {Szydagis}\ \emph {et~al.}(2011)\citenamefont
  {Szydagis}, \citenamefont {Barry}, \citenamefont {Kazkaz}, \citenamefont
  {Mock}, \citenamefont {Stolp}, \citenamefont {Sweany}, \citenamefont
  {Tripathi}, \citenamefont {Uvarov}, \citenamefont {Walsh},\ and\
  \citenamefont {Woods}}]{NEST:2011}%
  \BibitemOpen
  \bibfield  {author} {\bibinfo {author} {\bibfnamefont {M.}~\bibnamefont
  {Szydagis}}, \bibinfo {author} {\bibfnamefont {N.}~\bibnamefont {Barry}},
  \bibinfo {author} {\bibfnamefont {K.}~\bibnamefont {Kazkaz}}, \bibinfo
  {author} {\bibfnamefont {J.}~\bibnamefont {Mock}}, \bibinfo {author}
  {\bibfnamefont {D.}~\bibnamefont {Stolp}}, \bibinfo {author} {\bibfnamefont
  {M.}~\bibnamefont {Sweany}}, \bibinfo {author} {\bibfnamefont
  {M.}~\bibnamefont {Tripathi}}, \bibinfo {author} {\bibfnamefont
  {S.}~\bibnamefont {Uvarov}}, \bibinfo {author} {\bibfnamefont
  {N.}~\bibnamefont {Walsh}}, \ and\ \bibinfo {author} {\bibfnamefont
  {M.}~\bibnamefont {Woods}},\ }\href {\doibase 10.1088/1748-0221/6/10/P10002}
  {\bibfield  {journal} {\bibinfo  {journal} {Journal of Instrumentation}\
  }\textbf {\bibinfo {volume} {6}},\ \bibinfo {pages} {P10002} (\bibinfo {year}
  {2011})}\BibitemShut {NoStop}%
\bibitem [{\citenamefont {Wojcik}\ and\ \citenamefont
  {Tachiya}(2003)}]{Wojcik:2003}%
  \BibitemOpen
  \bibfield  {author} {\bibinfo {author} {\bibfnamefont {M.}~\bibnamefont
  {Wojcik}}\ and\ \bibinfo {author} {\bibfnamefont {M.}~\bibnamefont
  {Tachiya}},\ }\href {\doibase 10.1016/j.cplett.2003.08.006} {\bibfield
  {journal} {\bibinfo  {journal} {Chemical Physics Letters}\ }\textbf {\bibinfo
  {volume} {379}},\ \bibinfo {pages} {20} (\bibinfo {year} {2003})}\BibitemShut
  {NoStop}%
\bibitem [{\citenamefont {Wojcik}\ and\ \citenamefont
  {Tachiya}(2002)}]{Wojcik:2002}%
  \BibitemOpen
  \bibfield  {author} {\bibinfo {author} {\bibfnamefont {M.}~\bibnamefont
  {Wojcik}}\ and\ \bibinfo {author} {\bibfnamefont {M.}~\bibnamefont
  {Tachiya}},\ }\href {\doibase 10.1016/S0009-2614(02)01177-6} {\bibfield
  {journal} {\bibinfo  {journal} {Chemical Physics Letters}\ }\textbf {\bibinfo
  {volume} {363}},\ \bibinfo {pages} {381} (\bibinfo {year}
  {2002})}\BibitemShut {NoStop}%
\bibitem [{\citenamefont {Mozumder}(1995)}]{Mozumder:1995}%
  \BibitemOpen
  \bibfield  {author} {\bibinfo {author} {\bibfnamefont {A.}~\bibnamefont
  {Mozumder}},\ }\href {\doibase 10.1016/0009-2614(95)00384-3} {\bibfield
  {journal} {\bibinfo  {journal} {Chemical Physics Letters}\ }\textbf {\bibinfo
  {volume} {238}},\ \bibinfo {pages} {143} (\bibinfo {year}
  {1995})}\BibitemShut {NoStop}%
\bibitem [{\citenamefont {Sowada}, \citenamefont {Warman},\ and\ \citenamefont
  {de~Haas}(1982)}]{Sowada:1982}%
  \BibitemOpen
  \bibfield  {author} {\bibinfo {author} {\bibfnamefont {U.}~\bibnamefont
  {Sowada}}, \bibinfo {author} {\bibfnamefont {J.~M.}\ \bibnamefont {Warman}},
  \ and\ \bibinfo {author} {\bibfnamefont {M.~P.}\ \bibnamefont {de~Haas}},\
  }\href {\doibase 10.1103/PhysRevB.25.3434} {\bibfield  {journal} {\bibinfo
  {journal} {Physical Review B}\ }\textbf {\bibinfo {volume} {25}},\ \bibinfo
  {pages} {3434} (\bibinfo {year} {1982})}\BibitemShut {NoStop}%
\bibitem [{\citenamefont {Jaskolski}\ and\ \citenamefont
  {Wojcik}(2011)}]{Jaskolski:2011}%
  \BibitemOpen
  \bibfield  {author} {\bibinfo {author} {\bibfnamefont {M.}~\bibnamefont
  {Jaskolski}}\ and\ \bibinfo {author} {\bibfnamefont {M.}~\bibnamefont
  {Wojcik}},\ }\href {\doibase 10.1021/jp201149w} {\bibfield  {journal}
  {\bibinfo  {journal} {J. Phys. Chem. A}\ }\textbf {\bibinfo {volume} {115}},\
  \bibinfo {pages} {4317} (\bibinfo {year} {2011})}\BibitemShut {NoStop}%
\bibitem [{\citenamefont {{ICARUS Collaboration}}(2004)}]{Amoruso:2004}%
  \BibitemOpen
  \bibfield  {author} {\bibinfo {author} {\bibnamefont {{ICARUS
  Collaboration}}},\ }\href {\doibase 10.1016/j.nima.2003.11.423} {\bibfield
  {journal} {\bibinfo  {journal} {Nuclear Instruments and Methods in Physics
  Research Section A: Accelerators, Spectrometers, Detectors and Associated
  Equipment}\ }\textbf {\bibinfo {volume} {523}},\ \bibinfo {pages} {275}
  (\bibinfo {year} {2004})}\BibitemShut {NoStop}%
\bibitem [{\citenamefont {Jaskolski}\ and\ \citenamefont
  {Wojcik}(2009)}]{Jaskolski:2009}%
  \BibitemOpen
  \bibfield  {author} {\bibinfo {author} {\bibfnamefont {M.}~\bibnamefont
  {Jaskolski}}\ and\ \bibinfo {author} {\bibfnamefont {M.}~\bibnamefont
  {Wojcik}},\ }\href {\doibase 10.1007/s11164-009-0047-3} {\bibfield  {journal}
  {\bibinfo  {journal} {Research on Chemical Intermediates}\ }\textbf {\bibinfo
  {volume} {35}},\ \bibinfo {pages} {453} (\bibinfo {year} {2009})}\BibitemShut
  {NoStop}%
\bibitem [{\citenamefont {Henson}(1964)}]{Henson:1964}%
  \BibitemOpen
  \bibfield  {author} {\bibinfo {author} {\bibfnamefont {B.~L.}\ \bibnamefont
  {Henson}},\ }\href {\doibase 10.1103/PhysRev.135.A1002} {\bibfield  {journal}
  {\bibinfo  {journal} {Physical Review}\ }\textbf {\bibinfo {volume} {135}},\
  \bibinfo {pages} {A1002} (\bibinfo {year} {1964})}\BibitemShut {NoStop}%
\bibitem [{\citenamefont {Biagi}(1999)}]{Biagi:1999}%
  \BibitemOpen
  \bibfield  {author} {\bibinfo {author} {\bibfnamefont {S.}~\bibnamefont
  {Biagi}},\ }\href {\doibase 10.1016/S0168-9002(98)01233-9} {\bibfield
  {journal} {\bibinfo  {journal} {Nuclear Instruments and Methods in Physics
  Research Section A: Accelerators, Spectrometers, Detectors and Associated
  Equipment}\ }\textbf {\bibinfo {volume} {421}},\ \bibinfo {pages} {234}
  (\bibinfo {year} {1999})}\BibitemShut {NoStop}%
\bibitem [{\citenamefont {{Biagi-v8.9 Database}}()}]{lxcat:Biagi8.9}%
  \BibitemOpen
  \bibfield  {author} {\bibinfo {author} {\bibnamefont {{Biagi-v8.9
  Database}}},\ }\href@noop {} {}\bibinfo {howpublished}
  {\url{http://www.lxcat.laplace.univ-tlse.fr/}},\ \bibinfo {note} {retrieved
  Oct 8, 2012}\BibitemShut {NoStop}%
\bibitem [{\citenamefont {{Phelps Database}}()}]{lxcat:Phelps}%
  \BibitemOpen
  \bibfield  {author} {\bibinfo {author} {\bibnamefont {{Phelps Database}}},\
  }\href@noop {} {}\bibinfo {howpublished}
  {\url{http://www.lxcat.laplace.univ-tlse.fr/}},\ \bibinfo {note} {retrieved
  Oct 8, 2012}\BibitemShut {NoStop}%
\bibitem [{\citenamefont {Fink}\ and\ \citenamefont {Yates}(1970)}]{Fink:1970}%
  \BibitemOpen
  \bibfield  {author} {\bibinfo {author} {\bibfnamefont {M.}~\bibnamefont
  {Fink}}\ and\ \bibinfo {author} {\bibfnamefont {A.~C.}\ \bibnamefont
  {Yates}},\ }\href@noop {} {\bibfield  {journal} {\bibinfo  {journal} {Atomic
  Data and Nuclear Data Tables}\ }\textbf {\bibinfo {volume} {1}},\ \bibinfo
  {pages} {385} (\bibinfo {year} {1970})}\BibitemShut {NoStop}%
\bibitem [{\citenamefont {Riley}, \citenamefont {{MacCallum}},\ and\
  \citenamefont {Biggs}(1975)}]{Riley:1975}%
  \BibitemOpen
  \bibfield  {author} {\bibinfo {author} {\bibfnamefont {M.~E.}\ \bibnamefont
  {Riley}}, \bibinfo {author} {\bibfnamefont {C.~J.}\ \bibnamefont
  {{MacCallum}}}, \ and\ \bibinfo {author} {\bibfnamefont {F.}~\bibnamefont
  {Biggs}},\ }\href {\doibase 10.1016/0092-640X(75)90012-1} {\bibfield
  {journal} {\bibinfo  {journal} {Atomic Data and Nuclear Data Tables}\
  }\textbf {\bibinfo {volume} {15}},\ \bibinfo {pages} {443} (\bibinfo {year}
  {1975})}\BibitemShut {NoStop}%
\bibitem [{\citenamefont {Cullen}, \citenamefont {Perkins},\ and\ \citenamefont
  {Seltzer}(1991)}]{EEDL:1991}%
  \BibitemOpen
  \bibfield  {author} {\bibinfo {author} {\bibfnamefont {D.}~\bibnamefont
  {Cullen}}, \bibinfo {author} {\bibfnamefont {S.}~\bibnamefont {Perkins}}, \
  and\ \bibinfo {author} {\bibfnamefont {S.}~\bibnamefont {Seltzer}},\
  }\href@noop {} {\enquote {\bibinfo {title} {Tables and graphs of electron
  interaction cross 10 ev to 100 gev derived from the {LLNL} evaluated electron
  data library {(EEDL)}, z = 1 -- 100},}\ }\bibinfo {type} {Technical Report}\
  \bibinfo {number} {{UCRL-50400-Vol.31}}\ (\bibinfo  {institution} {Lawrence
  Livermore National Laboratory},\ \bibinfo {address} {Livermore, {CA}},\
  \bibinfo {year} {1991})\BibitemShut {NoStop}%
\bibitem [{\citenamefont {Perkins}\ \emph {et~al.}(1991)\citenamefont
  {Perkins}, \citenamefont {Cullen}, \citenamefont {Chen}, \citenamefont
  {Rathkopf}, \citenamefont {Scofield},\ and\ \citenamefont
  {Hubbell}}]{Cul1991}%
  \BibitemOpen
  \bibfield  {author} {\bibinfo {author} {\bibfnamefont {S.}~\bibnamefont
  {Perkins}}, \bibinfo {author} {\bibfnamefont {D.}~\bibnamefont {Cullen}},
  \bibinfo {author} {\bibfnamefont {M.}~\bibnamefont {Chen}}, \bibinfo {author}
  {\bibfnamefont {J.}~\bibnamefont {Rathkopf}}, \bibinfo {author}
  {\bibfnamefont {J.}~\bibnamefont {Scofield}}, \ and\ \bibinfo {author}
  {\bibfnamefont {J.}~\bibnamefont {Hubbell}},\ }\href@noop {} {\enquote
  {\bibinfo {title} {Tables and graphs of atomic subshell and relaxation data
  derived from the {LLNL} evaluated atomic data library ({EADL)}, z =
  1--100},}\ }\bibinfo {type} {Technical Report}\ \bibinfo {number}
  {{UCRL-50400-Vol.30}}\ (\bibinfo  {institution} {Lawrence Livermore National
  Laboratory},\ \bibinfo {address} {Livermore, {CA}},\ \bibinfo {year}
  {1991})\BibitemShut {NoStop}%
\bibitem [{Cul()}]{CulData:1991}%
  \BibitemOpen
  \href@noop {} {}\bibinfo {howpublished}
  {\url{http://www-nds.iaea.org/epdl97/}}\BibitemShut {NoStop}%
\bibitem [{\citenamefont {Barsanov}\ \emph {et~al.}(2007)\citenamefont
  {Barsanov}, \citenamefont {Dzhanelidze}, \citenamefont {Zlokazov},
  \citenamefont {Kotelnikov}, \citenamefont {Markov}, \citenamefont {Selin},
  \citenamefont {Shakirov}, \citenamefont {Abdurashitov}, \citenamefont
  {Veretenkin}, \citenamefont {Gavrin}, \citenamefont {Gorbachev},
  \citenamefont {Ibragimova}, \citenamefont {Kalikhov}, \citenamefont {Mirmov},
  \citenamefont {Shikhin}, \citenamefont {Yants}, \citenamefont {Khomyakov},\
  and\ \citenamefont {Cleveland}}]{Barsanov:2007}%
  \BibitemOpen
  \bibfield  {author} {\bibinfo {author} {\bibfnamefont {V.}~\bibnamefont
  {Barsanov}}, \bibinfo {author} {\bibfnamefont {A.}~\bibnamefont
  {Dzhanelidze}}, \bibinfo {author} {\bibfnamefont {S.}~\bibnamefont
  {Zlokazov}}, \bibinfo {author} {\bibfnamefont {N.}~\bibnamefont
  {Kotelnikov}}, \bibinfo {author} {\bibfnamefont {S.}~\bibnamefont {Markov}},
  \bibinfo {author} {\bibfnamefont {V.}~\bibnamefont {Selin}}, \bibinfo
  {author} {\bibfnamefont {Z.}~\bibnamefont {Shakirov}}, \bibinfo {author}
  {\bibfnamefont {D.}~\bibnamefont {Abdurashitov}}, \bibinfo {author}
  {\bibfnamefont {E.}~\bibnamefont {Veretenkin}}, \bibinfo {author}
  {\bibfnamefont {V.}~\bibnamefont {Gavrin}}, \bibinfo {author} {\bibfnamefont
  {V.}~\bibnamefont {Gorbachev}}, \bibinfo {author} {\bibfnamefont
  {T.}~\bibnamefont {Ibragimova}}, \bibinfo {author} {\bibfnamefont
  {A.}~\bibnamefont {Kalikhov}}, \bibinfo {author} {\bibfnamefont
  {I.}~\bibnamefont {Mirmov}}, \bibinfo {author} {\bibfnamefont
  {A.}~\bibnamefont {Shikhin}}, \bibinfo {author} {\bibfnamefont
  {V.}~\bibnamefont {Yants}}, \bibinfo {author} {\bibfnamefont
  {Y.}~\bibnamefont {Khomyakov}}, \ and\ \bibinfo {author} {\bibfnamefont
  {B.}~\bibnamefont {Cleveland}},\ }\href {\doibase 10.1134/S1063778807020111}
  {\bibfield  {journal} {\bibinfo  {journal} {Physics of Atomic Nuclei}\
  }\textbf {\bibinfo {volume} {70}},\ \bibinfo {pages} {300} (\bibinfo {year}
  {2007})}\BibitemShut {NoStop}%
\bibitem [{LBL()}]{LBL2007}%
  \BibitemOpen
  \href@noop {} {}\bibinfo {howpublished}
  {\url{http://ie.lbl.gov/toi/perchart.htm}}\BibitemShut {NoStop}%
\bibitem [{\citenamefont {Kazkaz}\ \emph {et~al.}(2010)\citenamefont {Kazkaz},
  \citenamefont {Foxe}, \citenamefont {Bernstein}, \citenamefont {Hagmann},
  \citenamefont {Jovanovic}, \citenamefont {Sorensen}, \citenamefont
  {Stoeffl},\ and\ \citenamefont {Winant}}]{Kazkaz:2010}%
  \BibitemOpen
  \bibfield  {author} {\bibinfo {author} {\bibfnamefont {K.}~\bibnamefont
  {Kazkaz}}, \bibinfo {author} {\bibfnamefont {M.}~\bibnamefont {Foxe}},
  \bibinfo {author} {\bibfnamefont {A.}~\bibnamefont {Bernstein}}, \bibinfo
  {author} {\bibfnamefont {C.}~\bibnamefont {Hagmann}}, \bibinfo {author}
  {\bibfnamefont {I.}~\bibnamefont {Jovanovic}}, \bibinfo {author}
  {\bibfnamefont {P.}~\bibnamefont {Sorensen}}, \bibinfo {author}
  {\bibfnamefont {W.}~\bibnamefont {Stoeffl}}, \ and\ \bibinfo {author}
  {\bibfnamefont {C.}~\bibnamefont {Winant}},\ }\href {\doibase
  10.1016/j.nima.2010.06.088} {\bibfield  {journal} {\bibinfo  {journal}
  {Nuclear Instruments and Methods in Physics Research Section A: Accelerators,
  Spectrometers, Detectors and Associated Equipment}\ }\textbf {\bibinfo
  {volume} {621}},\ \bibinfo {pages} {267} (\bibinfo {year}
  {2010})}\BibitemShut {NoStop}%
\bibitem [{\citenamefont {Yoshino}, \citenamefont {Sowada},\ and\ \citenamefont
  {Schmidt}(1976)}]{Yoshino:1976}%
  \BibitemOpen
  \bibfield  {author} {\bibinfo {author} {\bibfnamefont {K.}~\bibnamefont
  {Yoshino}}, \bibinfo {author} {\bibfnamefont {U.}~\bibnamefont {Sowada}}, \
  and\ \bibinfo {author} {\bibfnamefont {W.~F.}\ \bibnamefont {Schmidt}},\
  }\href {\doibase 10.1103/PhysRevA.14.438} {\bibfield  {journal} {\bibinfo
  {journal} {Physical Review A}\ }\textbf {\bibinfo {volume} {14}},\ \bibinfo
  {pages} {438} (\bibinfo {year} {1976})}\BibitemShut {NoStop}%
\bibitem [{\citenamefont {Huang}\ and\ \citenamefont
  {Freeman}(1981)}]{Huang:1981}%
  \BibitemOpen
  \bibfield  {author} {\bibinfo {author} {\bibfnamefont {S.~S.-S.}\
  \bibnamefont {Huang}}\ and\ \bibinfo {author} {\bibfnamefont {G.~R.}\
  \bibnamefont {Freeman}},\ }\href {\doibase 10.1103/PhysRevA.24.714}
  {\bibfield  {journal} {\bibinfo  {journal} {Physical Review A}\ }\textbf
  {\bibinfo {volume} {24}},\ \bibinfo {pages} {714} (\bibinfo {year}
  {1981})}\BibitemShut {NoStop}%
\end{thebibliography}
 
%

\end{document}